# Enhanced Superconductivity in TiO Epitaxial Thin Films

Chao Zhang[1], Feixiang Hao[1], Guanyin Gao[1], Xiang Liu[1], Chao Ma[1], Yue Lin[1], Yuewei Yin[*,1,2], and Xiaoguang Li[*,1,3,4]

[1]*Hefei National Laboratory for Physical Sciences at the Microscale, Department of Physics, University of Science and Technology of China, Hefei 230026, China*

[2]*Department of Physics and Astronomy, University of Nebraska, Lincoln, NE 68588, USA*

[3]*Key Laboratory of Materials Physics, Institute of Solid State Physics, CAS, Hefei 230026, China*

[4]*Collaborative Innovation Center of Advanced Microstructures, Nanjing 210093, China*

**Abstract**

Titanium oxides have many fascinating optical and electrical properties, such as the superconductivity at 2.0 K in cubic titanium monoxide TiO polycrystalline bulk. However, the lack of TiO single crystals or epitaxial films has prevented systematic investigations on its superconductivity. Here, we report the basic superconductivity characterizations of cubic TiO films epitaxially grown on (0001)-oriented $\alpha$-$Al_2O_3$ substrates. The magnetic and electronic transport measurements confirmed that TiO is a type-II superconductor and the recorded high $T_c$ is about 7.4 K. The lower critical field ($H_{c1}$) at 1.9 K, the extrapolated upper critical field $H_{c2}(0)$, and coherence length are about 18 Oe, 13.7 T and 4.9 nm, respectively. With increasing pressure, the value of $T_c$ shifts to lower temperature while the normal state resistivity increases. Our results on the superconducting TiO films confirm the strategy to achieve higher $T_c$ in epitaxial films, which may be helpful for finding more superconducting materials in various related systems.

---

[*] **Materials & Correspondence**: Correspondence and requests for materials should be addressed to X.G.L. (email: lixg@ustc.edu.cn) or to Y.W.Y. (email: yyin11@unl.edu). Fax/Tel: ++86-551-63603408.

**Introduction**

Transition metal oxides, such as titanium oxides, are a large family of materials with many fascinating electrical properties and applications[1-3]. Among various stable titanium oxides, the cubic metallic monoxide TiO is one of very interesting materials because of the extremely wide homogeneity range with oxygen content varying from about 0.80 to 1.30[4-7]. Electrical, optical, magnetic, and structural properties of bulk TiO have been widely investigated[8-17]. Especially, the superconductivity in TiO bulk materials was discovered in early 1965's by Hulm *et al*[18]. They reported that, for the bulk TiO with NaCl structure, its superconducting transition temperature $T_c$ increases from 0.2 K to 1.0 K with oxygen content increases from 0.9 to 1.1, but is below 0.08 K outside this range[19]. Further investigations showed that, high pressure annealing increases the oxygen content and lattice constant of cubic TiO bulk, and its superconducting transition temperature increases linearly with oxygen content to a maximum of 2.0 K[4,7]. However, the intrinsic superconducting properties of TiO, such as the lower and upper critical fields, superconducting coherence length and so on, are not clear yet, due to the difficulty in obtaining single crystals or epitaxial films. Even the zero resistance superconducting state has not been clearly reported. Therefore, high-quality TiO epitaxial thin films or single crystals are essential and highly desired for investigating the fundamental superconductivity of the system. There was only one ~10 nm TiO single crystalline film reported very recently, which was formed on $TiO_2$ substrate through surface chemical reduction method by a low-energy ion bombardment technique[20]. Although, this method is good at creating a thin TiO film on $TiO_2$ substrate, no magnetic and electrical properties concerning the superconductivity were reported. In fact, it is still a challenge to prepare high quality TiO epitaxial films on different substrates through a more controllable method like magnetron sputtering or pulsed laser deposition techniques.

On the other hand, it is well known that, for epitaxial superconducting thin films, the superconductivity could be enhanced or even created by proper heterostructure interfaces, such as the enhanced $T_c$ above 100 K in the epitaxial FeSe films grown on SrTiO$_3$ substrates, which was explained in terms of the coupling between conduction electrons and the substrate phonons[21,22,23], as well as the superconductivity created at the interface between two insulators like LaAlO$_3$/SrTiO$_3$[24,25]. In addition, a suitable lattice mismatch induced strain can also enhance the $T_c$ as reported in La$_{1.9}$Sr$_{0.1}$CuO$_4$[26]. Therefore, it will be interesting to see whether the superconductivity in TiO films can be improved as well.

In the present work, we successfully prepared cubic TiO thin films on a (0001)-oriented α-Al$_2$O$_3$ single crystal substrate by a pulsed laser deposition technique. The superconducting transition temperature of 7.4 K was observed for the TiO films, confirmed by magnetization and electrical transport measurements. The $T_c$ is almost 4 times higher than its bulk value and is suppressed with increasing pressure (Maximum pressure 1.8 GPa).

**Results and Discussions**

The TiO thin films with the thickness of ~80 nm were epitaxially grown on (0001)-oriented α-Al$_2$O$_3$ single crystalline substrates (see Methods). The structural characterization of the TiO thin films was performed using high angle annular dark-field scanning transmission electron microscope (HAADF-STEM) and *X*-ray diffraction (XRD). It was determined that the cubic TiO thin film grown on α-Al$_2$O$_3$ (0001) substrate is of the [111] direction perpendicular to its surface. Figure 1(a) shows the cross-sectional HAADF-STEM image at TiO/α-Al$_2$O$_3$ interface region of an as-grown sample viewed along the [1$\bar{1}$00] direction of sapphire, and thus the formation of a highly epitaxial TiO film is confirmed with a 2~3 atomic transition layer at the interface. The epitaxial relationship between the cubic TiO layer and sapphire substrate is determined as TiO [11$\bar{2}$] (111)//α-Al$_2$O$_3$ [11$\bar{2}$0] (0001). Fig. 1(b) shows the HAADF-STEM image and the related electron diffraction pattern in the TiO region viewed along the [1$\bar{1}$00] direction of sapphire as well, which shows single crystalline quality and agrees well with the simulation using the face-centered cubic (FCC) titanium monoxide structure (See supplementary information Fig. S1). Chemical and oxidation state analyses of the TiO film were performed by the electron energy-loss spectroscopy (EELS) in STEM, as shown in Fig. 1(c). For titanium oxides, the fine structures of the Ti-$L_{3,2}$ and O-$K$ edges reflect the covalent bonding states resulting from strong hybridization between Ti-3$d$ and O-2$p$ electronic states. For our sample, the Ti-$L_{3,2}$ and O-$K$ edges both consist mainly of two peaks which represent the fingerprint feature of cubic titanium monoxide (space group $Fm\bar{3}m$)[27]. The EELS spectra recorded from different areas of the TiO film indicate the structure homogeneity in the obtained TiO film, as shown in the supplementary information Fig. S2. Through the quantitative analysis of EELS results[28], it was found that, as shown in Fig. S2, the oxygen content in the film is not uniform and the averaged O/Ti ratio is about 1.11-1.25, which is similar to that reported by Pabon $et$ $al$[20].

Although the combination of STEM images with electron diffraction pattern is a powerful tool in determining the crystal structure of materials, the results provide only local structural information of a small volume. Therefore, we carried out detailed XRD experiments. Fig. 1(d) shows the XRD $\theta/2\theta$ specular scan of an as-grown TiO thin film on $\alpha$-Al$_2$O$_3$ (0001) substrate. Only the (111) family diffractions of TiO film and the sapphire (0006) diffraction peak were observed, indicating that the film is highly orientated in [111] direction parallel to sapphire [0001]. The cubic lattice constant about 4.164 Å of the TiO film is calculated from the peak position of (111) plane using Bragg's law, which is very close to the titanium monoxide bulk value of 4.177 Å. One of the reasons for the difference in lattice parameters between the TiO film and the bulk may be related to the oxygen content, because it was reported that the lattice parameter could be reduced with increasing oxygen content[7,19,20]. As shown in the inset of Fig. 1(d), the full width at the half maximum (FWHM) of the TiO (111) plane rocking curve is about 0.3° indicating its good crystallinity. To obtain the in-plane epitaxial relationship between the TiO film and sapphire substrate more clearly, XRD $\phi$-scans of the (200) plane of the TiO film and the ($10\bar{1}4$) plane of $\alpha$-Al$_2$O$_3$ substrate were performed, as shown in Fig. 1(e). For the TiO film, the six distinct peaks at 60° intervals to each other with nearly the same intensity indicate a six-fold rotational symmetry along the TiO (111) plane normal. From the XRD $\phi$-scan, an epitaxial relation as TiO [$11\bar{2}$] (111)//$\alpha$-Al$_2$O$_3$ [$11\bar{2}0$] (0001) was obtained, consistent with the STEM results shown in Fig. 1(a-b).

The existence of the superconducting phase in the TiO film was unambiguously confirmed by its Meissner effect and zero resistance. Fig. 2(a) shows the direct-current (DC) magnetizations *versus* temperature (*M-T*) of the TiO film. Both the field-cooled (FC) and zero-field-cooled (ZFC) magnetizations in 20 Oe magnetic field perpendicular to the film surface indicate the appearance of superconductivity near 7.0 K, much higher than that of TiO polycrystalline bulk (0.2 K ~ 2.0 K)[4,19]. The irreversible region of magnetizations marked by the bifurcation of FC and ZFC curves below $T_c$ shows smaller FC signal as compared with the ZFC signal, indicating the flux trapping by thin film defects during the field-cooled process. The magnetization *versus* magnetic field curve *M-H* at 1.9 K for magnetic field perpendicular to the film surface ($H\perp(111)$) is shown in Fig. 2(b). The lower critical field $H_{c1}$, defined as the field at which a flux first penetrates, can be estimated from the *M-H* curve as a deviation from the linear *M-H* behavior corresponding to the Meissner state. The evaluated value is about 18 Oe at 1.9 K for $H\perp(111)$ as shown in the inset of Fig. 2(b).

The superconducting property of the TiO film was also investigated through electrical transport measurements using the Hall bar pattern. The schematic diagram of the experimental setup is shown in the supplementary information Fig. S3. Fig. 3 shows the temperature dependent resistance ($R_{xx}$) in zero magnetic field and Hall resistance ($R_{xy}$) in 2 T magnetic field. In the $R_{xx}$-$T$ curve, the resistance rises with decreasing temperature first, and a kink ($T_{kink}$) appears around 130 K below which the resistance increases even steeper. This kink may be related to the charge localization[29,30]. When the temperature goes down further, the resistance decreases suddenly and the superconducting behavior occurs at ~7.4 K, as shown in the inset of Fig. 3(a). The superconducting resistive transition at zero field is relatively broad, which may be attributed to the following two aspects: i) inhomogeneous oxygen or titanium stoichiometry of the films, as shown in Fig. S2, and ii) a consequence of a Berezinskii–Kosterlitz–Thouless (BKT)-like transition, as described in the following equation[31]:

$$R_{xy}(T) = R_N \exp[-b(T/T_{BKT}-1)^{(-1/2)}] \qquad (1)$$

where $R_N$ is normal-state sheet resistance, and $b$ is a non-universal dimensionless numerical constant. The good fitting result (blue solid line in inset of Fig. 3(a)) indicates that the resistive transition may be a consequence of the BKT-like transition ($T_{BKT}$ = 4.91 K) probably related to the TiO film -$Al_2O_3$ substrate 2D interface which localizes and enhances the superconductivity through conduction electron-substrate phonon interaction, similar to that occurred in FeSe thin films [22,23].

From the Hall result, it is confirmed that the *n*-type electronic charge carriers dominate the conduction mechanism, as shown in Fig. 3(b). According to the relation $R_H = R_{xy}d/H = 1/nq$, where $R_H$, $d$, $n$, and $q$ are the Hall coefficient, film thickness, charge carrier density, and carrier charge, respectively, the carrier concentration was estimated to be about 2.0 × $10^{22}$ cm$^{-3}$ at 300 K. It is noted that the Hall voltage near the superconducting transition temperature has sign opposite to the voltage in the normal state, as shown in the inset of Fig. 3(b), which may be related to the unusual vortex motion[32,33].

Fig. 4(a-b) show the temperature dependent resistances with different magnetic fields parallel ($H$//(111)) and perpendicular ($H\perp$(111)) to the TiO film surface. The parallel resistance broadenings in different fields are clearly observed, and the superconducting transition shifts to lower temperatures with increasing magnetic field. Fig. 4(c) shows the upper critical field $H_{c2}(T)$ and irreversibility field $H_{irr}(T)$ of the TiO film in the in-plane and out-of-plane configurations, determined using the criterions of 90% and 0.1% normal-state resistance (see supplementary Fig. S4 for the detailed information)[34,35]. The upper critical field *versus* temperature curves can be well fitted by[36]:

$$H_{c2}(T) = H_{c2}(0)[1-(T/T_c)^2] \quad (2)$$

here, $H_{c2}(0)$ is the upper critical field at absolute zero temperature. The extrapolated values of $H_{c2}$ are about 13.6 T and 13.7 T for magnetic fields perpendicular ($H\perp$(111)) and parallel ($H$//(111)) to the TiO film surface, respectively. Correspondingly, the Landau–Ginzburg superconducting coherence lengths, $\xi = [(h/2e)/(2\pi H_{c2})]^{1/2}$, at absolute zero temperature are estimated to be about 4.92 nm and 4.91 nm for $H\perp$(111) and $H$//(111), respectively. According to the anisotropic effective mass Ginzburg-Landau theory[37], the anisotropy ratio $\varepsilon$ ~ 1.01 was obtained from the scaling law $\varepsilon = (m_\perp/m_{//})^{0.5} = H_{c2}^{//}/H_{c2}^{\perp}$. Here $H_{c2}^{\perp}$ and $H_{c2}^{//}$ are the upper critical fields for $H\perp$(111) and $H$//(111), and $m_\perp$ and $m_{//}$ are the effective masses of electrons along these directions, respectively. As for the temperature dependence of $H_{irr}$, it can be well fitted by[36]:

$$H_{irr}(T) = H_{irr}(0)(1-T/T_c)^n \quad (3)$$

with $n$=0.85. To investigate the anisotropic superconducting properties in more detail, we studied the magnetic field orientation dependence of the superconducting transition. Fig. 4(d) shows the field dependences of the resistances at different angles $\beta$ at 4.0 K, where $\beta$ denotes the tilt angle between the normal of film plane and the field direction, as depicted in the inset of Fig. 4(d). One can see that $H_{c2}$ changes a little, while $H_{irr}$ increases gradually with increasing $\beta$ from $H\perp(111)$ to $H//(111)$.

Fig. 5 shows the pressure effect on the superconducting transition of the TiO film. With increasing pressure, the zero-resistance transition temperature shifts to lower temperatures and the normal state resistivity increases obviously. It may be due to the pressure enhanced charge localization, and the effective attraction of the Cooper pairs is thus suppressed as a result.

From the above results and discussions, it is clear that a significant $T_c$ enhancement was observed in the TiO epitaxial thin film. Although the underlying physics mechanism is not well understood, our results suggest several clues for the occurrence of superconductivity. First, inhomogeneous oxygen or titanium stoichiometry was detected from the detailed STEM analysis, which may significantly affect the local electronic structure as well as the superconductivity in the cubic TiO film. Second, the BKT-like transition in the TiO thin films, similar to the FeSe thin films[23], may be probably related to the TiO-$Al_2O_3$ interface through the conduction electron-substrate phonon interaction[22]. Third, the high pressure experiments demonstrate the importance of strain effect on the superconductivity. The strain status in the TiO film can be influenced by not only the substrate strain but also the oxygen stoichiometry. Further experimental and theoretical investigations on the micro-stoichiometry, structure, and electronic properties both in the non-uniform films and at the TiO-$Al_2O_3$ interface are required to finalize the mechanism of the enhanced $T_c$.

In conclusion, the superconducting cubic TiO thin films were epitaxially grown on the (0001)-oriented $\alpha$-$Al_2O_3$ single crystalline substrates, and the superconducting properties of the films were systematically characterized. It was found that the superconducting transition temperature is enhanced to about 7.4 K, and the upper critical field at zero field $H_{c2}(0)$ is about 13.7 T ($H//(111)$) in the TiO thin films. The high pressure experimental results indicated that increasing pressure weakens the superconductivity. Our work will facilitate prospects for understanding the superconducting mechanism in the titanium-based oxide superconductors as well as achieving higher temperature superconductivity.

**Methods**

**Sample preparation**. The TiO thin films with the thickness of about 80 nm were epitaxially grown on commercial (0001)-oriented $\alpha$-$Al_2O_3$ single crystalline substrates by a pulsed laser deposition technique. Prior to the film deposition, the substrates were annealed in flowing oxygen at 1100 °C for 120 min. A KrF excimer laser ($\lambda$ = 248 nm) was employed to ablate the $Ti_2O_3$ target. The chamber was evacuated to a base pressure of ~ $1\times10^{-7}$ Torr and purged three times with high purity nitrogen gas. Then, the films were deposited at 900°C and cooled down to room temperature naturally in high vacuum. The laser energy density, repetition rate, and target-substrate distance used for the deposition were 3.5–4.0 J·cm$^{-2}$, 5 Hz, and 4.5 cm, respectively.

**Structural characterizations.** High-resolution XRD measurements were performed with a commercial Panalytical X'pert x-ray diffract meter with the Cu K$\alpha_1$ radiation at wavelength of 1.5406 Å. For the structural and chemical characterization of the TiO/$Al_2O_3$ samples, a JEOL JEM-ARM200F operating at 200 kV, equipped with a spherical aberration corrector on the condenser lens system, was used to obtain the high angle annular dark field (HAADF) scanning transmission electron microscopy (STEM) images and core-level electron energy-loss spectroscopy (EELS) spectra. The thickness and surface roughness of the TiO films were measured by STEM and atomic force microscope scanning probe microscope (AFM MultiMode V), as shown in the supplementary information Fig. S5.

**Hall bar fabrication.** Ohmic contact electrodes of gold/titanium films with a Hall bar geometry were fabricated by ultraviolet light engraving machine JKG-2A and ion beam sputtering coating system DPS-LIM, which enabled measurements of the four-terminal resistance and Hall coefficient of the channel. A schematic illustration is shown in the supplementary information Fig. S3. The dimensions of the channel were 500 μm in width and 3 mm in length. The resistance and the Hall coefficient were measured in a Physical Property Measurement System (PPMS-14, Quantum Design).

**Electrical and magnetic measurements**. Magnetization measurements were carried out using a Squid Vibrating Sample Magnetometer (Squid-VSM, Quantum Design). The angle resolved in-plane four-probe resistance measurements were performed for the TiO films rotated from out-of-plane to in-plane of the film in magnetic fields up to 14 T in a Physical Property Measurement System (PPMS-14, Quantum Design). The angle $\beta = 0°$ is defined as the applied magnetic field was perpendicular to surface of the TiO film ($H\perp(111)$).

**High-pressure experiments.** The pressure dependences of resistances of the TiO film were measured in a Pressure Cell of Physical Property Measurement System (PPMS-9, Quantum Design). The value of the pressure was demarcated by $T_c$ of Sn metal ($T_c$ = 3.72 K at atmospheric pressure for Sn).

**Acknowledgments**

This work was supported by the Natural Science Foundation of China (51332007, 21521001 and 51622209) and by the National Basic Research Program of China (2015CB921201, 2016YFA0300103 and 2012CB922003).

**Competing interests**

The authors declare no competing financial interests.

**Contributions**

X.G.L. and Y.W.Y. designed and supervised the experiments. C.Z. and H.F.X. fabricated the sample and performed electric transport, SEM and AFM measurements. G.Y.G. carried out XRD measurements. X.L. and C.Z. performed electronic transport measurements under high pressure. C.M., Y.L. and C.Z. carried out STEM-HAADF and EELS measurements and analyzed the relevant data; X.G.L., C.Z. and H.F.X. were responsible for all data analysis and wrote the manuscript. All the authors contributed to the discussions and editing of the manuscript.

**Figure legends**

**Figure 1 | Structural and chemical characterizations.** (a-b) HAADF-STEM images of $TiO/Al_2O_3$, at (a) interface region and (b) TiO region. (c) Ti-$L_{3,2}$ and O-$K$ edges EELS spectra of the TiO film. (d) $\theta/2\theta$ XRD pattern of the TiO/$\alpha$-$Al_2O_3$(0001) heterostructure. The inset shows the rocking curve for the TiO (111) reflection. (e) XRD $\phi$ scan performed on TiO (200) and sapphire ($10\bar{1}4$) planes.

**Figure 2 | Magnetization characterizations.** (a) Temperature dependence of the DC magnetization of TiO film in ZFC and FC modes at 20 Oe for $H\perp(111)$. The $T_c$ is identified as 7.0 K. (b) Magnetization *vs.* magnetic field (*M-H*) at 1.9 K for $H\perp(111)$. Inset: $H_{c1}$ at 1.9 K was estimated to be about 18 Oe.

**Figure 3 | Electrical transport measurements.** (a) Temperature dependence of the resistance $R_{xx}$ from 1.9 K to 300 K. (b) Temperature dependence of Hall resistance $R_{xy}$ at a magnetic field of 2 T. The insets are the corresponding magnified figures around $T_c$ and the blue solid line is the fitting result by BKT-like transition.

**Figure 4 | Upper critical field and irreversibility field.** (a-b) Temperature dependent resistances in different magnetic fields (a) parallel ($H//(111)$) and (b) perpendicular

($H\perp(111)$) to the TiO film surface. (c) Temperature dependent $H_{c2}$ and $H_{irr}$ in different magnetic field directions $H\perp(111)$ and $H//(111)$. The hollow triangles and solid dots are experimental results deduced from (a) and (b), and the lines are the fitting results. (d) Magnetic field dependencies of the resistances at 4.0 K with different angles $\beta$ (defined as the inset). The inset shows magnetic field angle dependences of $H_{irr}$ and $H_{c2}$.

**Figure 5 | Pressure dependence of zero-resistance transition temperature.** Inset: $R_{xx}$-$T$ curves at different pressures.

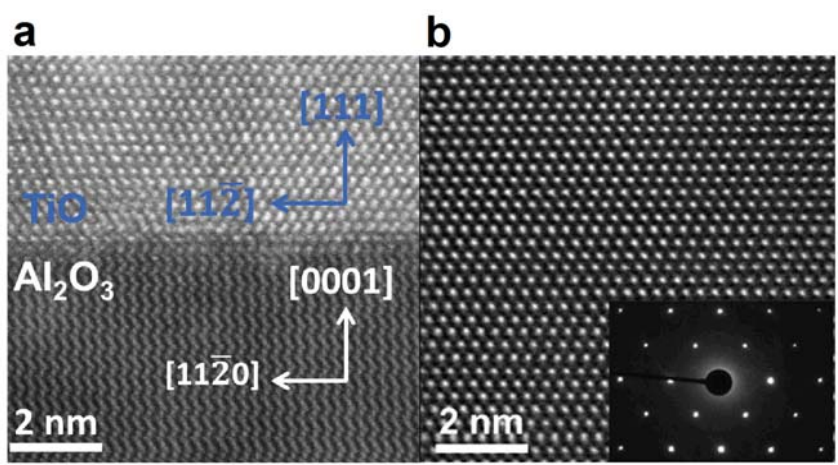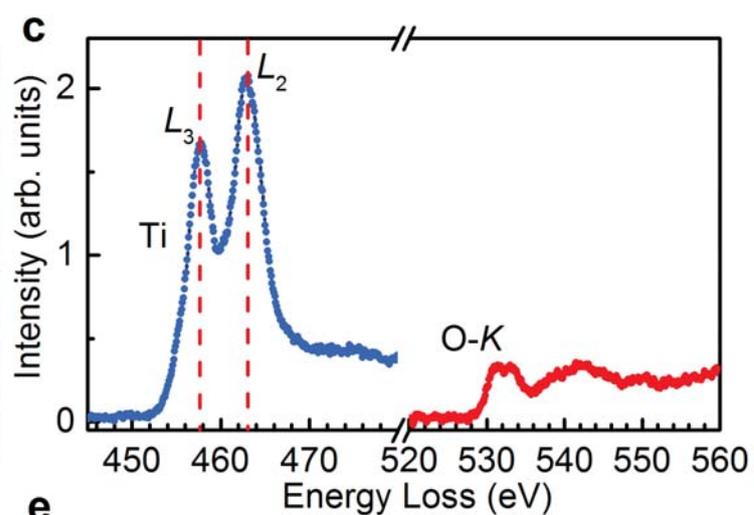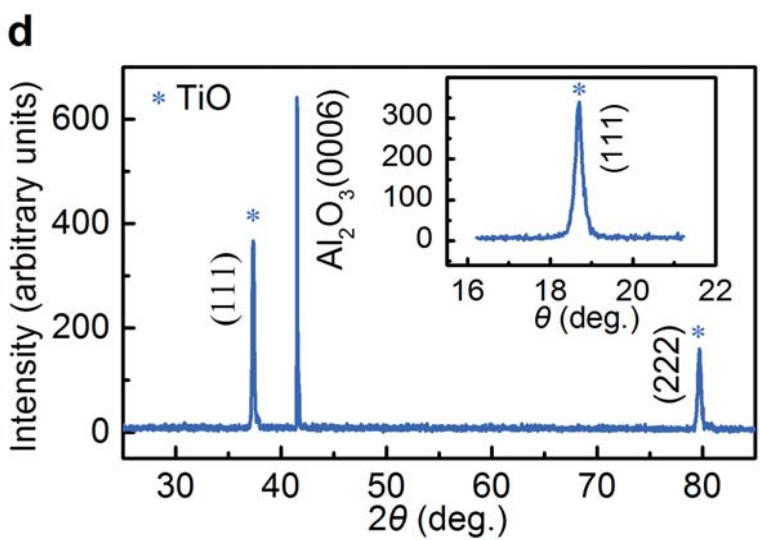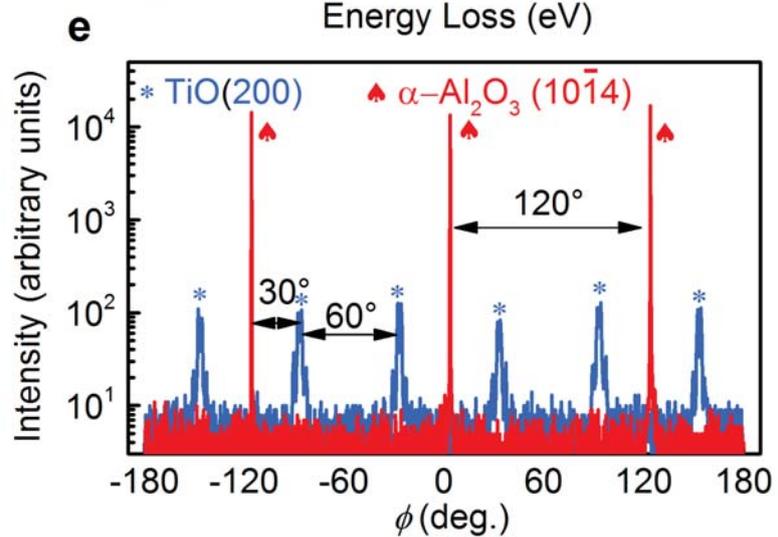

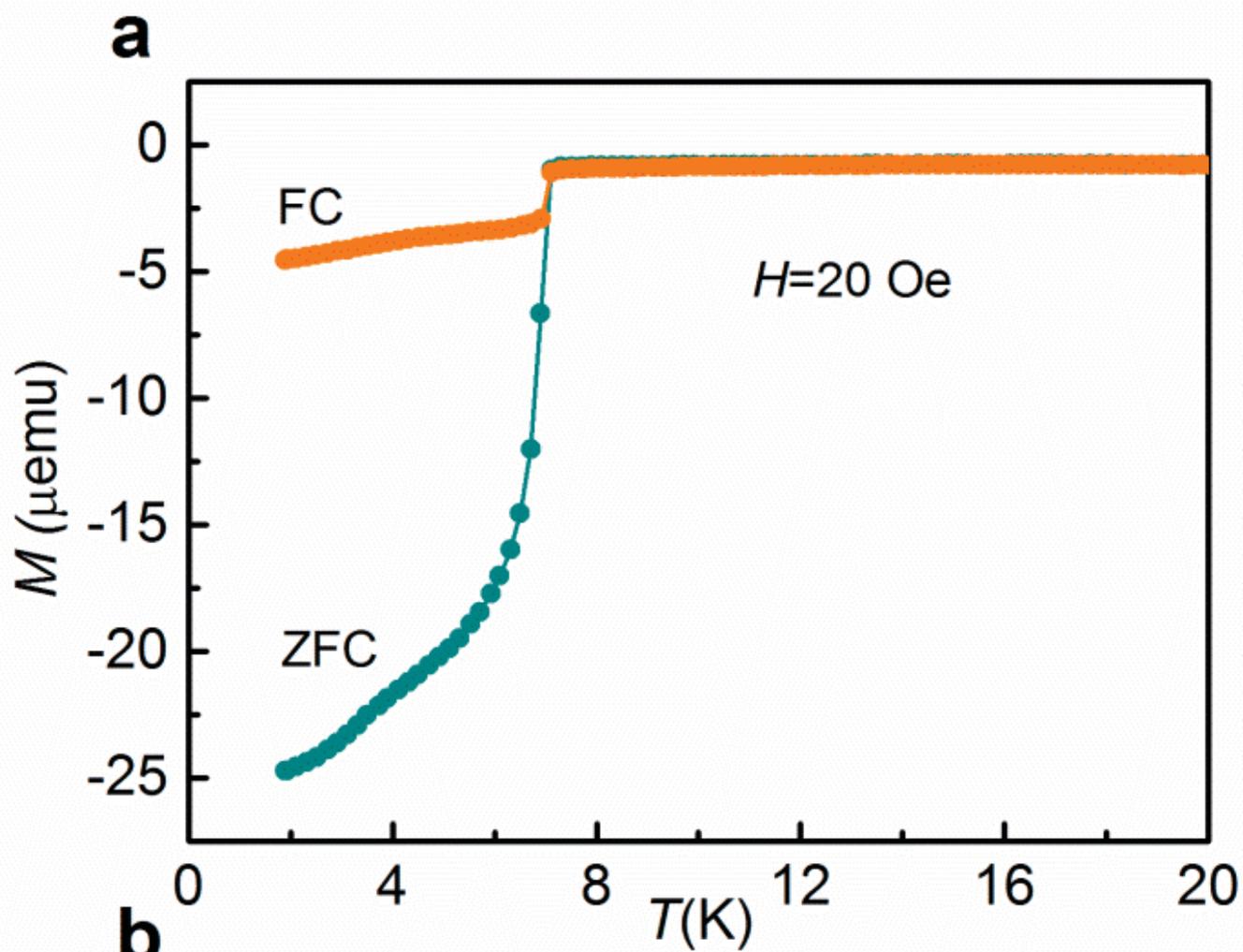
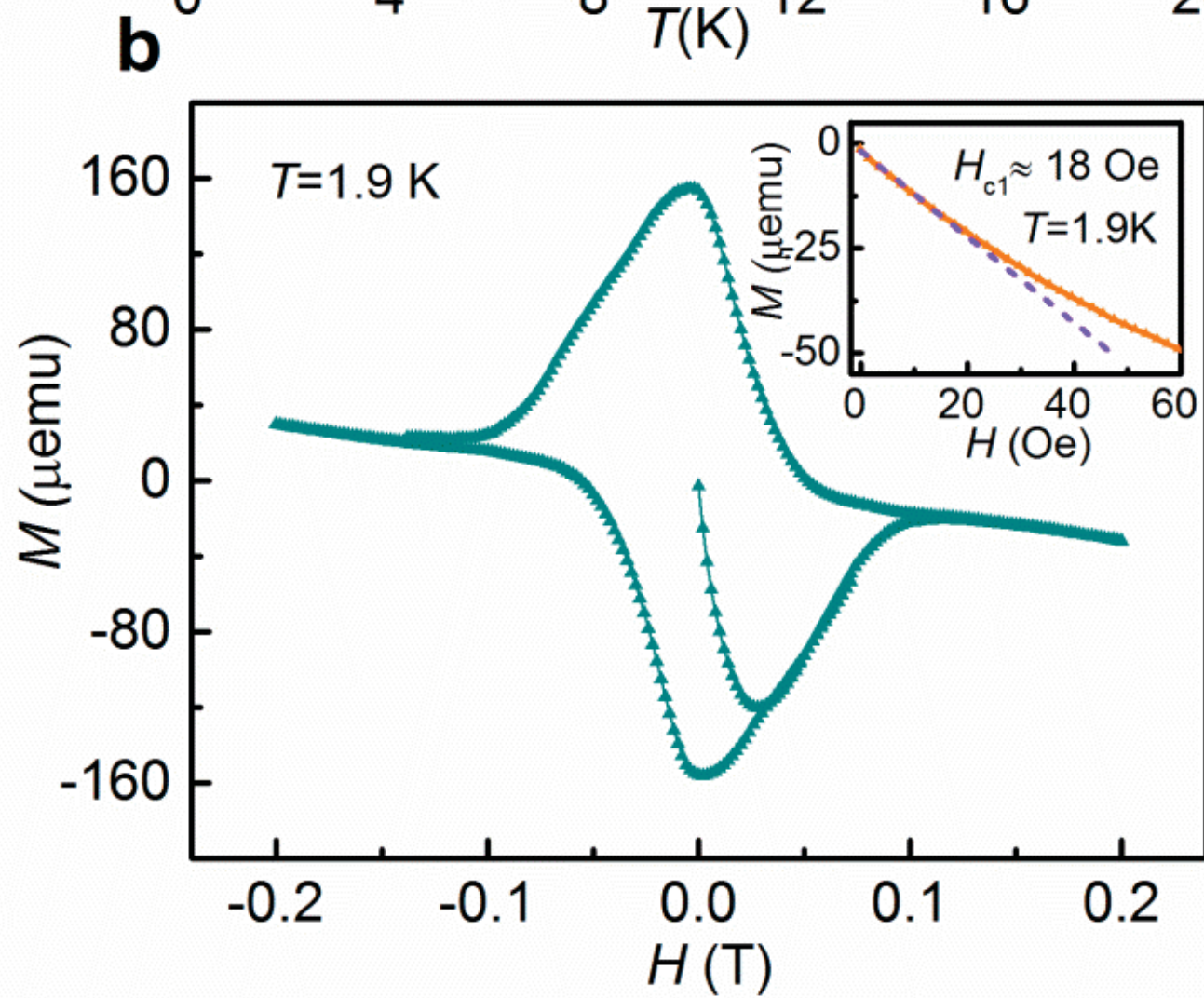

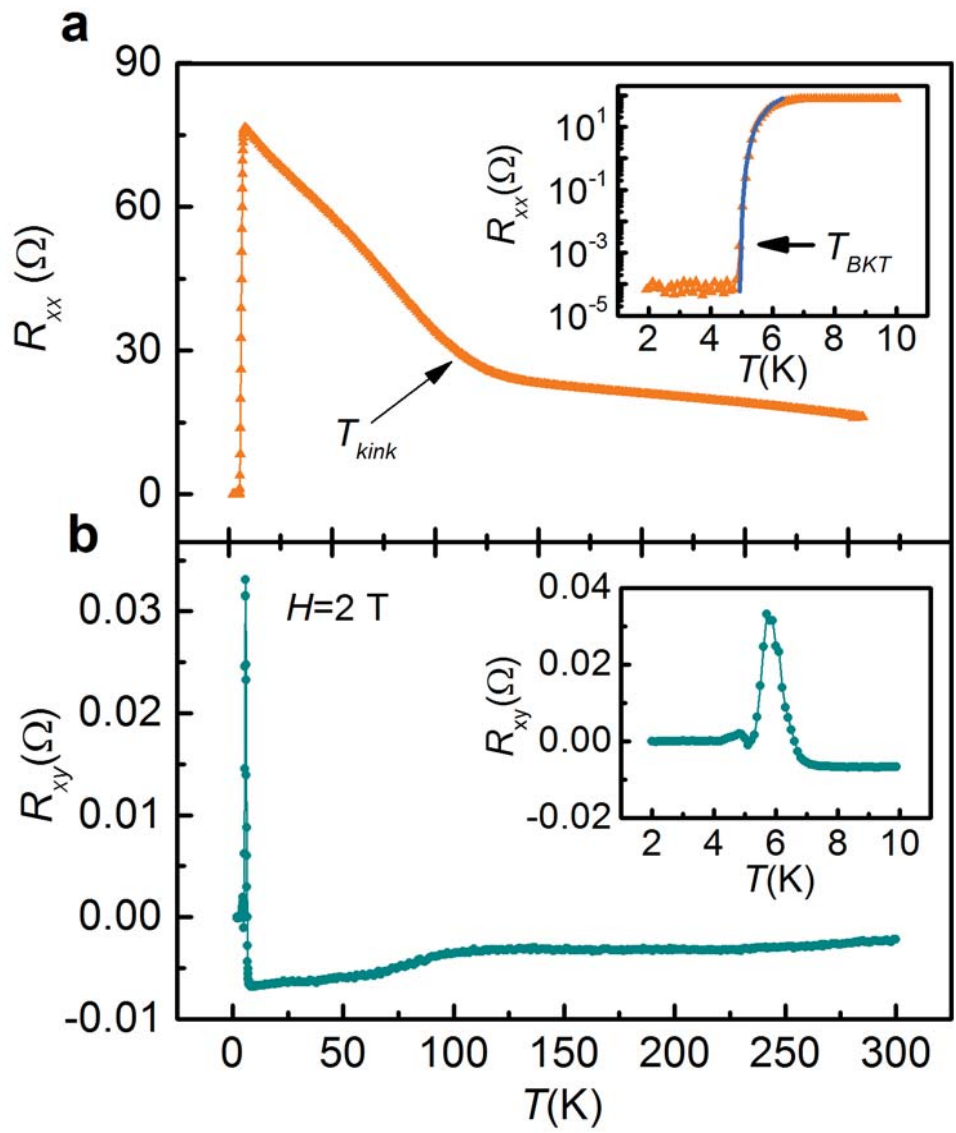

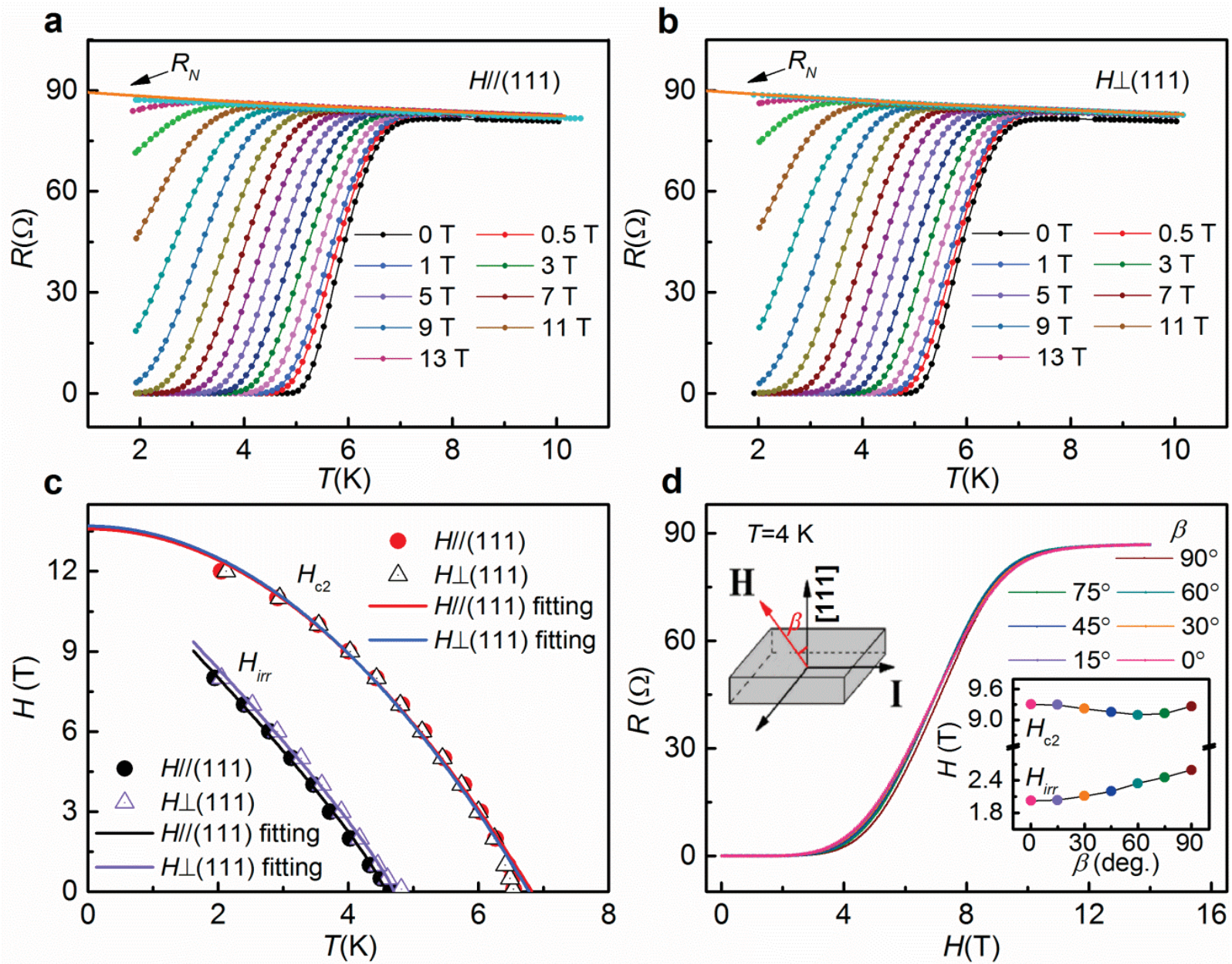

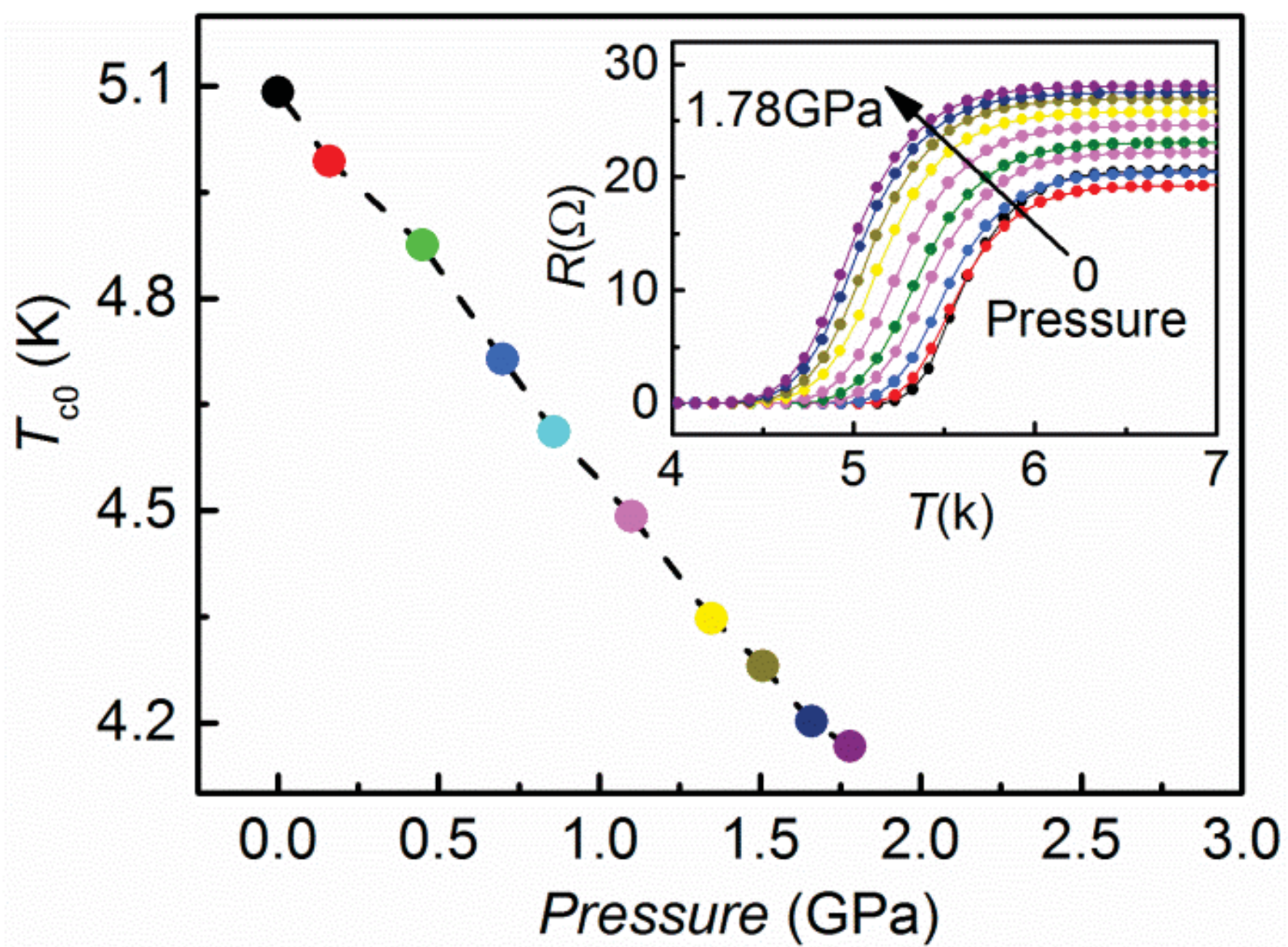

# Enhanced Superconductivity in TiO Epitaxial Thin Films

Chao Zhang[1], Feixiang Hao[1], Guanyin Gao[1], Xiang Liu[1], Chao Ma[1], Yue Lin[1], Yuewei Yin[*1,2], and Xiaoguang Li[*1, 3, 4]

[1]Hefei National Laboratory for Physical Sciences at the Microscale, Department of Physics, University of Science and Technology of China, Hefei 230026, China

[2]Department of Physics and Astronomy, University of Nebraska, Lincoln, NE 68588, USA

[3]Key Laboratory of Materials Physics, Institute of Solid State Physics, CAS, Hefei 230026, China

[4]Collaborative Innovation Center of Advanced Microstructures, Nanjing 210093, China

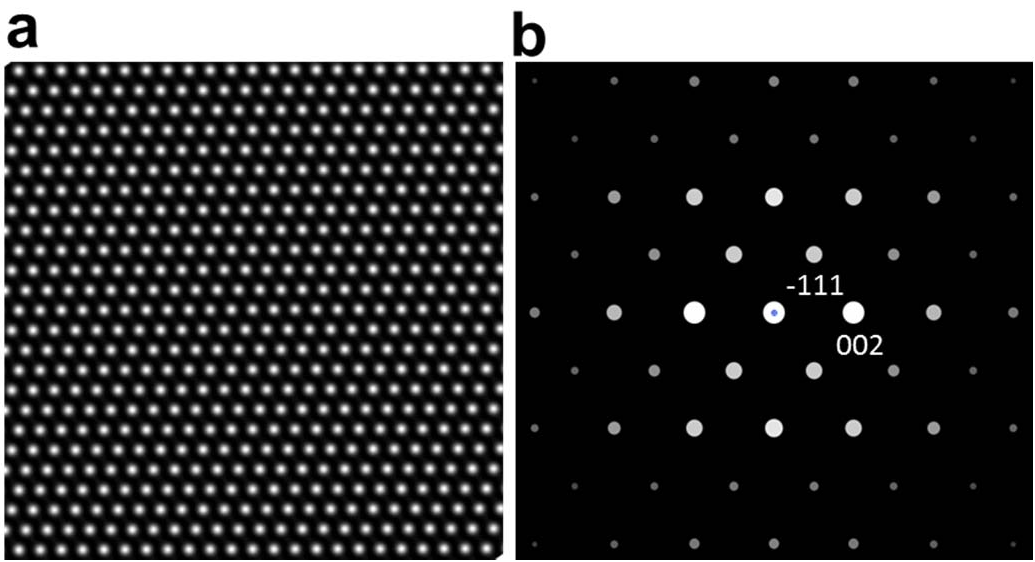

**Figure S1** Simulated HAADF-STEM image and the corresponding electron diffraction pattern of TiO viewed along the [110] direction using the face-centered cubic (FCC) titanium monoxide structure, which shows good consistency with experimental data shown in Fig. 1(b).

---

[*] Corresponding Authors: lixg@ustc.edu.cn, or yyin11@unl.edu

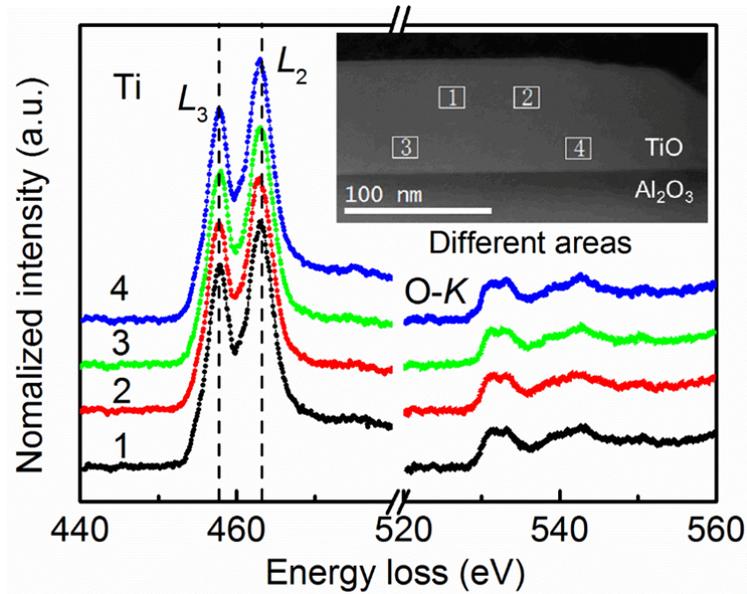

**Figure S2** The Ti-$L_{3,2}$ and O-$K$ EELS spectra of the TiO film at different positions. Identical EELS profiles were observed. Through the quantitative analysis of EELS results[1], it was found that the oxygen content in the film is not uniform and that the O/Ti ratios are ~1.18, ~1.25, ~1.11 and ~1.15 at different areas 1, 2, 3 and 4, respectively. The oxygen content is similar to that reported by Pabon *et al*[2].

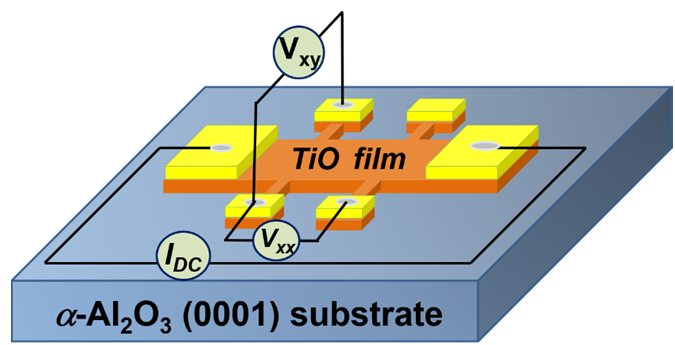

**Figure S3** Schematic diagram of the experimental setup for electrical transport measurements.

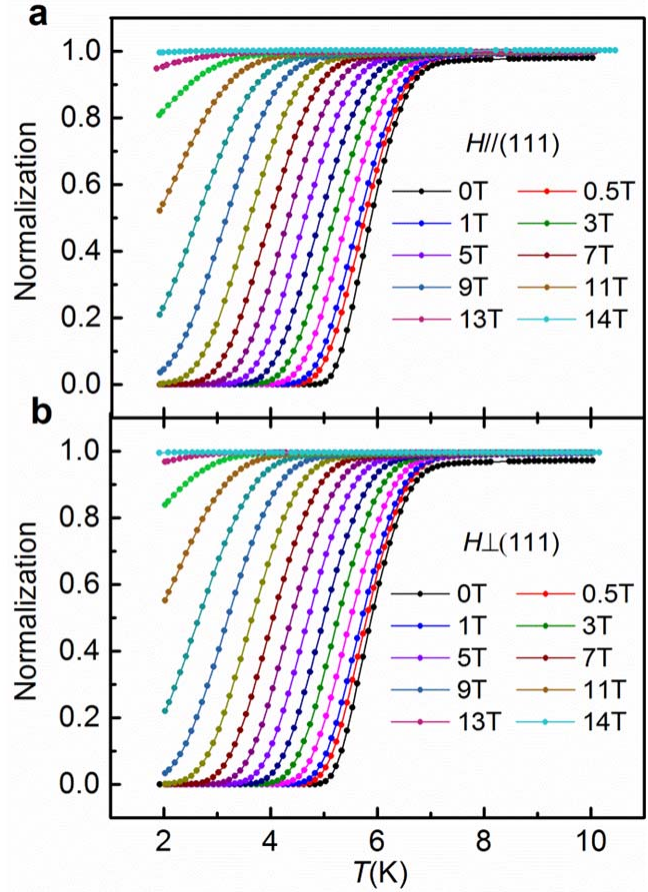

**Figure S4** Temperature dependent resistances normalized to its normal state value $R_N(T)$ in different magnetic fields parallel ($H//(111)$) and perpendicular ($H\perp(111)$) to TiO film surface. The current is applied along the film surface and perpendicular to the magnetic fields. The parallel resistance broadenings in different fields are clearly observed, and the superconducting transition shifts to lower temperature with increasing magnetic fields. The upper critical field $H_{c2}(T)$ and irreversibility field $H_{irr}(T)$ of in-plane and out-of-plane of TiO film are determined using the criterions of 90% and 0.1% normal-state resistivity normalized to its normal state value $R_N(T)$ [3,4]. Here, the magnetic field independent $R_N(T)$ is extrapolated from $R(T)$ curve at temperatures well above the transition temperature by a polynomial function.

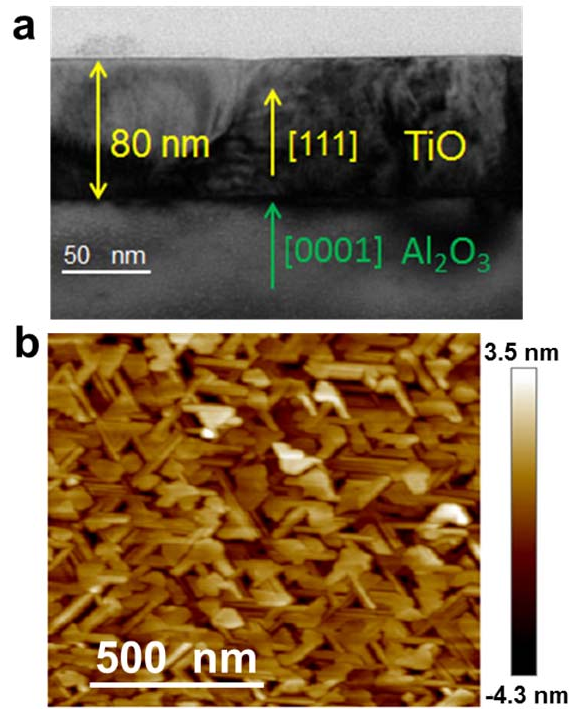

**Figure S5** (a) The cross-sectional TEM image indicates that the thickness of the TiO film is about 80 nm. (b) The atomic force microscope (AFM) image. The RMS roughness is about 1.01 nm.